\documentstyle{mn}
\newif\ifAMStwofonts

\ifoldfss
  \ifCUPmtlplainloaded \else
    \NewTextAlphabet{textbfit} {cmbxti10} {}
    \NewTextAlphabet{textbfss} {cmssbx10} {}
    \NewMathAlphabet{mathbfit} {cmbxti10} {} 
    \NewMathAlphabet{mathbfss} {cmssbx10} {} 
  \fi
  \ifAMStwofonts
    \ifCUPmtlplainloaded \else
      \NewSymbolFont{upmath} {eurm10}
      \NewSymbolFont{AMSa} {msam10}
      \NewMathSymbol{\upi}     {0}{upmath}{19}
      \NewMathSymbol{\umu}     {0}{upmath}{16}
      \NewMathSymbol{\upartial}{0}{upmath}{40}
      \NewMathSymbol{\leqslant}{3}{AMSa}{36}
      \NewMathSymbol{\geqslant}{3}{AMSa}{3E}

    \fi
  \fi
\fi 

\ifnfssone
  \newmathalphabet{\mathit}
  \addtoversion{normal}{\mathit}{cmr}{m}{it}
  \addtoversion{bold}{\mathit}{cmr}{bx}{it}
  \newmathalphabet{\mathbfit} 
  \addtoversion{normal}{\mathbfit}{cmr}{bx}{it}
  \addtoversion{bold}{\mathbfit}{cmr}{bx}{it}
  \newmathalphabet{\mathbfss} 
  \addtoversion{normal}{\mathbfss}{cmss}{bx}{n}
  \addtoversion{bold}{\mathbfss}{cmss}{bx}{n}
  \ifAMStwofonts
    \ifCUPmtlplainloaded \else
      \UseAMStwoboldmath
      \makeatletter
      \new@mathgroup\upmath@group
      \define@mathgroup\mv@normal\upmath@group{eur}{m}{n}
      \define@mathgroup\mv@bold\upmath@group{eur}{b}{n}
      \edef\UPM{\hexnumber\upmath@group}
      \new@mathgroup\amsa@group
      \define@mathgroup\mv@normal\amsa@group{msa}{m}{n}
      \define@mathgroup\mv@bold\amsa@group{msa}{m}{n}
      \edef\AMSa{\hexnumber\amsa@group}
      \makeatother
      \mathchardef\upi="0\UPM19
      \mathchardef\umu="0\UPM16
      \mathchardef\upartial="0\UPM40
      \mathchardef\leqslant="3\AMSa36
      \mathchardef\geqslant="3\AMSa3E
    \fi
  \fi
\fi 

\ifnfsstwo
  \DeclareMathAlphabet{\mathbfit}{OT1}{cmr}{bx}{it}
  \SetMathAlphabet\mathbfit{bold}{OT1}{cmr}{bx}{it}
  \DeclareMathAlphabet{\mathbfss}{OT1}{cmss}{bx}{n}
  \SetMathAlphabet\mathbfss{bold}{OT1}{cmss}{bx}{n}
  \ifAMStwofonts
    \ifCUPmtlplainloaded \else
      \DeclareSymbolFont{UPM}{U}{eur}{m}{n}
      \SetSymbolFont{UPM}{bold}{U}{eur}{b}{n}
      \DeclareSymbolFont{AMSa}{U}{msa}{m}{n}
      \DeclareMathSymbol{\upi}{0}{UPM}{"19}
      \DeclareMathSymbol{\umu}{0}{UPM}{"16}
      \DeclareMathSymbol{\upartial}{0}{UPM}{"40}
      \DeclareMathSymbol{\leqslant}{3}{AMSa}{"36}
      \DeclareMathSymbol{\geqslant}{3}{AMSa}{"3E}
    \fi
  \fi
\fi 

\ifCUPmtlplainloaded \else
  \ifAMStwofonts \else 
    \def\upi{\pi}
    \def\umu{\mu}
    \def\upartial{\partial}
  \fi
\fi

\title[UV and X-ray Absorption in PG1126$-$041]
{Ionized Ultraviolet and Soft-X-ray Absorption in the Low Redshift Active 
Galactic Nucleus PG1126$-$041}
\author[T. Wang et al.]
{T.G.~Wang,$^{1,2}$ W.~Brinkmann,$^3$  W.~Wamsteker$^4$, W.~Yuan$^3$ 
\newauthor
and J.X. Wang$^1$\\
$^1$ Center for Astrophysics, University of Science and Technology of China, 
Hefei,
230026 Anhui, China.\\
$^2$ The Institute of Physical and Chemical Research (RIKEN), 2-1, Hirosawa, 
Wako, Saitama, 351-0198, Japan\\
$^3$ Max-Planck-Institut f\"ur extraterrestrische Physik, Postfach 1603, D-85740
Garching, Germany.\\
$^4$ ESA IUE Observatory, P.O.Box 50727, 28080 Madrid, Spain.}
\date{Accepted *****. Received *****}
\pagerange{\pageref{firstpage}---\pageref{lastpage}}
\pubyear{1999}

\def\LaTeX{L\kern-.36em\raise.3ex\hbox{a}\kern-.15em
    T\kern-.1667em\lower.7ex\hbox{E}\kern-.125emX}

\begin{document}

\label{firstpage}

\maketitle

\begin{abstract}

We present here the analysis of ultraviolet spectra from IUE and an X-ray 
spectrum from {\it ROSAT} PSPC observations of the X-ray weak, 
far-infrared loud AGN, PG 1126$-$041 (Mrk 1298). 
The first UV spectra taken in June 1992, simultaneously with ROSAT, show 
strong absorption lines of NV, CIV and SiIV, extending 
over a velocity range from $-1000$ to  $-5000$ km~s$^{-1}$ with respect to the 
corresponding line centre. Our analysis shows that the Broad Emission Line 
Region (BELR) is, at least partially, covered by the material causing these 
absorption lines.
In the IUE spectrum taken in Jan. 1995, the continuum was a factor of two brighter and the UV absorption lines are found to be considerably weaker than in 
1992, but only little variation in the emission line fluxes is found.   
With UV spectral indices  of $\alpha_{uv}\simeq$ 1.82 and 1.46 for the 1992 and 
1995 data, the far UV spectrum is steep.  Based on the emission line ratios and 
the broad band spectral energy distribution, we argue that the steepness 
of the UV spectrum is unlikely to be due to reddening.

The soft X-ray emission in the ROSAT band is weak. A simple power-law 
model yields a very poor fit with a  UV-to-X-ray spectral index 
$\alpha_{uvx}=2.32$.  Highly ionized (warm) absorption  
is suggested by the {\it ROSAT} data. After correcting for a warm absorber, 
the optical to X-ray spectral slope is close to the average of  
$\alpha_{uvx}\simeq 1.67$ for radio quiet quasars. 

From photoionization calculations we find:  (1) A single zone 
absorption model cannot explain simultaneously the UV absorption lines and 
the ionized X-ray absorption if metal abundances are solar. Furthermore, 
in order to be consistent with the equivalent width of the observed $Ly\alpha$ 
absorption line, the turbulent velocity of the warm absorber must be less than 
190 km s$^{-1}$, which imposes serious constraints on a disk wind model.  
(2) The UV absorption lines and their variability cannot be explained by a 
single zone model with solar abundances and the large variability in the 
absorption lines suggest that CIV and NV absorption lines are not 
severely saturated. (3) The absorption of the ionizing continuum by warm 
material strongly affects the emission line spectrum. 
\end{abstract}


\section{Introduction}

About 10\% of radio quiet luminous quasars (QSOs) exhibit Broad Absorption Lines (BAL) in
the highly ionized resonant UV lines, such as NV, SiIV and CIV (Turnshek 1984,
 Weymann, 1995). Approximately  5--15\% of these also show low ionization BAL 
 such as MgII, AlIII (Weymann et al. 1991, Hartig \& Baldwin 1990
, Boroson \& Meyers 1992, Voit, Weyman \& Korista 1993). The BALs are 
displaced to shorter wavelengths with velocities of up to 0.1c relative to their 
corresponding emission line peaks. 
This indicates the presence of partially ionized
 gas outflows at enormous velocities from the central nucleus in these objects.
The BAL QSOs are either a unique type of objects or normal QSOs
viewed under special geometrical conditions.
 Evidence in  support of  the latter interpretation includes the
small covering factor inferred from the line profile modeling (Hamann et al.
1993) and the similarity in the emission line spectra between BAL and
non-BAL objects (Weymann et al. 1991). New spectro-polarimetric 
observations reveal that the BAL region shows a disk-like structure and that the BAL 
clouds are mixed with dust which, at least in some objects, effectively blocks 
part of the light from the innermost nucleus  (Glenn, Schmidt \& Foltz 1994, 
Goodrich \& Miller 1995,Cohen et al. 1995, Hines \& Wills 1995).

Blue-shifted, low velocity absorption lines are usually found in the IUE spectra 
of $\sim$15\% of less luminous, low redshift, Seyfert galaxies. Variations of 
these absorption lines imply that they are also produced close
to the nucleus, as are the BAL clouds (Stocke et al. 1994  and references 
therein). These absorption lines might represent the low velocity counterpart of the 
BAL phenomenon in the less luminous Seyfert nuclei.
 Stocke et al. (1994) proposed that the low velocities are the consequence of the comparatively low total power of Seyfert galaxies.

The UV spectrum of BAL QSOs has been extensively studied, while the X-ray spectra of 
these objects -potentially important to clarify a number of questions- have been 
poorly investigated. The unresolved absorption troughs in the UV may contain 
partially 
saturated lines, which can easily escape the detection in a moderate resolution 
spectrum. This is not a problem for the X-ray continuum, because the X-ray 
absorption is sensitive to the total absorbing column density in the BAL region.
Part of the absorbing material might be highly ionized and produces only
very weak absorption in the UV but much stronger absorption in X-rays.

The total hydrogen column density inferred from the UV absorption lines is of
the order $\sim$10$^{20}$~cm$^{-2}$, which is not high enough to produce
significant X-ray absorption, however there exists some evidence for large X-ray
absorption in BAL objects. The X-ray emission of BAL objects in the {\it ROSAT}
band is generally extremely weak. Only very few objects have been  detected in soft X-rays
(Green \& Mathur 1996, Wang, Brinkmann \& Bergeron 1996), with a broad band spectral index 
of  $\alpha_{ox}>1.9$  for all of them. This suggests that the X-ray emission of
BAL QSOs is either intrinsically weak or suppressed by absorption.
The similarity in the emission line spectra of  BAL and non-BAL QSOs, and the 
detection of large X-ray absorption in two relative bright BAL QSOs (Green \& 
Mathur 1996, Mathur, Elvis \& Singh 1996 ) suggest that absorption is 
common in these objects. 
Some recent BAL models predict the presence of significant  X-ray absorption. To 
overcome the confinement problem (c.f.  Weymann 1995) in radiative   
acceleration models for BAL QSO's,  Murray et al. (1995) suggested that the BAL 
gas
is shielded by a very large column density (10$^{23}$--10$^{24}$~cm$^{-2}$) of 
highly ionized gas,  which absorbs the soft X-ray photons but not the 
ultraviolet photons. Murray \& Chiang (1996) further extended these results
to Seyfert galaxies and radio-loud objects. Absorbing  column densities of
a few times 10$^{22}$~cm$^{-2}$ are predicted for Seyfert galaxies and 
radio-loud quasars. Although this model explains nicely the soft X-ray deficit in BAL 
QSOs, verification of this theory is largely dependent on the detection of the 
warm absorbing gas.
The existence of thick absorber has been claimed for a few BAL QSOs, it has 
however not been demonstrated that the gas is highly ionized, as is essential for the validity  of this theory.
It should be noted that most low redshift Seyfert galaxies with UV 
absorption lines do not show heavily absorbed X-ray spectra, i.e., steep optical 
to X-ray spectrum, although about 50\% of these objects do show some warm 
absorber associated features (Reynolds 1996).

In this paper, we report the result of the detailed analysis of the ultraviolet 
and X-ray spectrum of the high redshift BAL QSO analogue, PG1126$-$041 (Mrk 
1298) which is a high luminosity Seyfert I galaxy ($M_v=-22.8$, only 
slightly fainter than a typical quasar) with weak X-ray emission 
($\alpha_{ox}=2.05$). Strong UV absorption, extending to a maximum velocity of 
about $-5000$ km~s$^{-1}$,  in CIV, NV and SiIV are present in the low resolution IUE spectra. The {\it ROSAT} PSPC spectrum shows strongly ionized absorption with a column density $\simeq 4\times 10^{22}$~cm$^{-2}$.

\section{ The Ultraviolet spectrum}

Two short-wavelength and one long 
wavelength spectrum were obtained with IUE in June 1992 and one short wavelength 
spectrum in January 1995. The spectra were retrieved from the IUE archive and 
were processed with the IUE NEWSIPS software, which also corrects thermal and 
temporal degradation of the camera sensitivity. They have been corrected for 
Galactic reddening, which has been estimated from the Galactic Hydrogen column 
density 
$4.4\times 10^{20}$ ~cm$^{-2}$ (Dickey \& Lockman 1990) with a conversion
factor $5.5\times 10^{21}$~cm$^{-2}$ (Diplas \& Savage 1993). All line and 
continuum measurements reported in this paper are based on the spectra corrected for the corresponding reddening of E(B-V) = 0.08. 

In Fig. 1 we show the combined SWP and LWP spectra of 1992. Emission and 
absorption lines are marked in the figure. The strong NV, CIV and SiIV absorption lines, at the short wavelength side of the emission lines, are pronounced. The short wavelength spectrum taken in 1995 (SWP53285) shows an increase in the continuum flux by a factor two (see Table~1), much weaker absorption lines (Table~2), and little variation in the emission lines. 

Figure 2 shows details of the emission and absorption line profiles at both 
epochs. The 1992-spectrum is normalized to match the local continuum of 
the 1995 data for each line.  Firstly, we notice that CIV emission line peaks to 
the red side of the line position in the object's rest 
frame (z=0.060), unlike the Ly$\alpha$ line, which  peaks at zero velocity. This 
difference is most likely due to the strong absorption in CIV. Secondly, in the 1992 
spectrum, the CIV and NV absorption troughs are deeper than the local continuum, 
suggesting partial coverage of the absorbing material of the emission line 
region. In 1995, the absorption lines are much weaker and an 
additional large velocity CIV absorption line is indicated.
 
As the emission lines are relatively narrow, the continuum, which can normally 
only be defined with difficulty in BAL QSOs, can be reliably estimated from the 
pseudo-line free windows.
The continuum over the full SWP region is modeled with a single power
law. Emission lines are fitted by multiple Gaussians:
 Ly$\alpha$~ and CIV are modeled with two Gaussians with the relative central 
wavelengths and widths for each component fixed;
NV 1240, SiIV+OIV], HeII, CIII] and MgII are fitted with one Gaussian and
their central wavelengths are fixed at the observed wavelengths.

Due to the low spectral resolution of IUE, the detailed structure of the 
absorption lines can hardly be resolved. Therefore we modeled each broad 
absorption line with one Gaussian, applied to both the continuum and the
emission line. 

The best fit parameters are derived by minimizing  $\chi^2$~, taking into
account the NEWSIPS errors. The fit is done using the SPECFIT package
developed by G. Kriss within the distribution of STSDAS.
 Finally, the whole wavelength range of the SWP camera from 1200\AA~ to 
1930\AA~ is simultaneously fitted, allowing a reliable determination of both 
the continuum shape and the flux.
The initial values were estimated by fitting each line separately over a 150\AA~ 
window centered on the observed emission line peak. Only two strong absorption 
lines, NV and CIV, were included  in this first step. The weaker absorption 
lines  in SiIV and Ly$\alpha$, are then added  after a reasonable fit is 
achieved. The centre and width of the absorption lines in velocity space are 
locked to those of the CIV absorption line.  To avoid finding a local minimum, a 
'simplex' search is used throughout the fitting process.

Since a single power law  continuum cannot be defined over the full LWP band, 
due to 
contamination by UV FeII and Balmer continuum emission, we applied only local 
continuum 
fits, for CIII] and MgII, over a window width of 300 \AA~ centred on the 
observed wavelength of each line. The fit over the profile of the MgII line 
yields a line width 2430 km~s$^{-1}$, similar to 
that of the H$\beta$ line, allowing for the separation of the doublet nature of 
MgII and the resolution of LWP camera. No MgII absorption line is visible.

The final fit over the SWP band is shown in Fig 3.
The best fit parameters for the continuum are  listed in Table~1  for the 
emission lines and the continuum, and in Table~2  for the absorption 
lines. The emission line flux is the sum of the individual components.  
The errors given in Table~1 and Table~2 are purely statistical 
at the 1$\sigma$~ level.
Since there exists no evidence for short time-scale UV variations in Seyfert 
galaxies of comparable luminosity, the difference between the two spectra taken 
in 1992 can be regarded as an indicator for the errors  introduced by
the measurements and  the calibration. The central velocity ( $-1900$ km$^{-1}$ )
and width (3000 km~s$^{-1}$ ) for the absorption lines from the different  
measurements are in good agreement. The equivalent widths of NV, SiIV, and CIV 
absorption lines as 
well as the fluxes of the Ly$\alpha$, NV, SiIV+OIV], and CIV emission lines from 
the two spectra in 1992 are also within their statistical errors. The HeII 
fluxes differ 
by a factor of two, but the HeII line is badly blended with OIII]$\lambda$1663, 
and forms a very broad feature, and the line flux depends critically on the
exact value of the continuum placement. In the 1992 spectra a Ly$\alpha$~ 
absorption line is required for spectrum SWP44823, but not for the other (SWP 44822).

Assuming optically thin conditions, column densities for the absorbing ions can 
be estimated from the above equivalent widths: $W_\lambda/\lambda~=~\pi 
e^2/(m_e~c^2)~Ng\lambda f~\simeq~8.85~\times~10^{-13} Ng\lambda f$, where 
$\lambda$~ and $f$~ are  the wavelength and oscillator strength of the 
corresponding absorption lines, {\it N} is the column density of the absorbing 
ion and $g$~ is the effective statistical weight for the ground level of the
corresponding ion (Spitzer 1978). $(g\lambda f)$ values were taken from
Korista et al. (1991). If individual optically thick absorption lines 
are present or the absorbing material only partially covers the continuum or 
emission line region, the column densities derived represent lower limits. These column densities are given in table~2 under N$_{1992}$.

The overall continuum in the SWP band is steep with a spectral index
$\alpha\simeq~1.5-1.8$ ($F_\nu~\propto~\nu^{-\alpha}$) similar to that seen in 
some low ionization BAL QSOs.  If the intrinsic UV continuum of PG1126$-$041 is 
similar to other QSOs (Francis et al. 1991), significant intrinsic reddening is 
required (see the discussion for PG0043+039 by Turnshek et al. 1994). However, 
the cold $N_H$ column density derived from the X-ray analysis cannot provide 
sufficient reddening if the dust-to-gas ratio and the properties of the dust are 
Galactic. Moreover, the Galactic 2200\AA~ absorption feature which is usually 
seen in similarly reddened objects, is weak. 
  
\section {The ROSAT X-ray Spectrum}

The ROSAT observation of PG 1126-041 was made overlapping in time with the IUE 
observation in 1992 with an exposure time of $\simeq$ 20 ks, and can be 
considered to be simultaneous. The methods used in the X-ray spectral analysis, 
including source and background extraction, dead-time and vignetting correction, 
and spectral fitting are similar to those described in Wang, Brinkmann \& 
Bergeron (1996). Independent analysis using EXSAS (WY) and XSELECT plus 
XSPEC(TW) gives essentially the same results.  The results presented here in 
Table~3, are those obtained from XSELECT. All errors quoted, are at 2.7$\sigma$ 
level for each single parameter of interest.

Although the net counts for the source are only about 320, a single power law
with Galactic column fails to produce a reasonable fit ($\chi^2_\nu$/d.o.f.
~=~4.00/10) leaving a deep and broad dip around 0.8 keV (Figure 4). 
In addition, this fit yields a column density ($N_H \simeq 0.5\times 
10^{20}$~cm$^{-2}$), some 10 times less than the Galactic value, and a very flat 
spectrum $\Gamma_x\simeq 1.48$. When the column density is fixed at the Galactic 
value, the fit becomes much worse with $\chi^2_\nu/d.o.f. ~=~4.25/11$.  In this 
case the single power law description the UV to X-ray spectral index is 
2.3$\pm$0.1, much steeper than for typical radio quiet QSOs with a mean value 
 1.65 (Yuan et al. 1998).  

As the photon deficit around 0.8-1.0 keV  is a typical signature of warm 
absorption due to edges, more complicated models were applied. As a first step, 
a single absorption edge was added to the model. A good fit can be obtained with 
($\chi^2_\nu/\nu=1.01/8$, Table 3). However, the edge energy of the best fit at
0.56$\pm$0.03 keV does not correspond to any of the more common ion edges. This 
might be caused by the combination of several edges. To evaluate the results 
further, we next fitted the spectrum with a warm absorption model using a variable 
slope power-law ionizing continuum (see Zdziarski et al. 1995). The free 
parameters are photon index, column density, absorber temperature and ionization 
parameter ($\xi=L/nR^2$).  We have fixed the temperature at $5\times 10^4$~K and 
forced the photon index of ionizing continuum to be the same as the X-ray photon 
index. The best fit parameters are presented in table 3. 
The fit is accepted at a confidence level of 23\% ($\chi^2_\nu/\nu=1.23/8$, 
Figure 4). A small amount of 
excess cold absorption is also required with $N_c = 2.6_{-0.9}^{+1.0}\times 
10^{20}$~cm$^{-2}$. Figure 5 shows $\xi$ versus $N_w$ contours for the warm 
absorption model. 
 The best fit column density $N_w=3.2_{-0.8}^{+0.9}\times 10^{22}$~cm$^{-2}$ 
for the absorbing material is within the range typically for Seyfert I galaxies 
(
e.g., Reynolds 1997).  However, the ionization parameter is lower than
normally found for other Seyfert I galaxies. Using the best fit 
photon index, the X-ray derived dimensionless ionization parameter 
$U_x =${\it (density of ionizing photons in the energy $>$ 0.1 keV)/(hydrogen 
density)} 
(Netzer 1996) is only 0.086$_{-0.010}^{+0.026}$ with $\xi = 55_{-12}^{+16}$.  
After correcting for the warm absorption, the far UV to X-ray spectral slope
becomes 1.66$_{-0.10}^{+0.07}$, which is consistent with the X-ray spectral 
index $1.79_{-0.19}^{+0.13}$ and well within the range of the mean value for 
radio quiet AGNs.  

\section{Discussion}

\subsection {Broad Band Continuum and Absorption}

In figure 6 we show the Spectral Energy Distribution (SED) of 
PG 1126-041 from infrared to X-ray energies. The SED peaks around 3000\AA~ and 
is flat in the infrared and optical band. As mentioned in the previous section, 
the far UV spectrum is steep and consistent with a direct  extrapolation of the 
X-ray spectrum. Barvainis (1993) interpreted the 
flatness of the infrared to optical spectrum in this object as contamination of 
stellar light which fills the gap between the spectral bumps. However, the steep 
UV spectrum cannot be explained in this way since the contribution of stellar 
light in the far UV is negligible under any reasonable assumption for the 
stellar population. An alternative method to generate such steep UV spectra is 
through significant dust absorption. We will show below that this is also an 
unlikely cause for the steep spectrum. 

If the intrinsic UV spectrum of PG1126-041 is similar to other QSOs a reddening 
of E(B-V) $>$0.15 is required. However, the Balmer decrement in this 
object is normal, with a ratio H$\alpha$/H$\beta$ = 2.92 (Miller et al. 1992), 
very close to that expected 
for case B recombination, and it is also similar to  the mean value of 
H$\alpha$/H$\beta$  = 3.07$\pm$0.56 as determined for a sample of bright QSOs 
with Z$<0.5$ by Miller et al. (1992). Using the H$\alpha$ flux of 8.92~10$^{-
13}$~erg~cm$^{-2}$~s$^{-1}$, we find from the $Ly\alpha$ flux in Table 1, that for 
PG1126-041 the ratio $Ly\alpha/H\alpha$ is between 2.3 and 3.6, for the two 
epochs of IUE observations in 1992 and 1995. As this is already very close to 
the photoionization prediction of $Ly\alpha/H\alpha$ $\simeq$4.0 , it is clear 
that reddening can not bring the flux at 1200\AA~ up much more than a factor of 
two at most, arguing against the existence of large reddening affecting the 
emission lines. 
 
As any absorbed UV light must be re-emitted in the infrared band, the 
infrared luminosity due to the dust emission must be a factor of two larger 
than the observed luminosity in the UV if the dust covers a large fraction of the 
nucleus and the reddening is as large as E(B-V)=0.15. The observed integrated 
infrared flux in the 1-100$\mu m$ band is 5.6$\times10^{-11}$~erg~cm$^{-2}$~s$^{-
1}$, which is similar to the integrated UV flux from 3000\AA~to 100\AA~ of 
5.7$\times10^{-11}$~erg~cm$^{-2}$~s$^{-1}$ for the 1992 data. In addition   
a significant portion of infrared emission has to originate from the host 
galaxy. 
Finally, the shortage of soft X-ray photons would even be more severe if the UV 
spectrum were highly reddened. With a E(B-V)=0.15 correction applied to the UV 
spectrum, the UV flux will increase by a factor of four, bringing  $\alpha_{uvx}$ 
back to $>2.0$, making the object intrinsically very weak in X-rays. 
Also, the absence of a strong 2200\AA~ feature implies that the grains must be 
different from the standard Galactic composition.

Although we can not completely rule out the possibility of that reddening is 
responsible for the steep UV spectrum, it requires a number of rather 
restrictive constraints: 1) the dust covers only a small  fraction of the BLR; 
(2) the grains are not of Galactic composition, (3) the X-ray emission is 
intrinsically weak. On the other hand, the UV spectrum of an object with UV 
absorption lines could very well be intrinsically steep. This can be associated 
for example, with inclination effects. The case of an intrinsically steep spectrum will 
be discussed in the next section.

\subsection {On the Ultraviolet and Ionized X-ray Absorption}

We have shown above that, in addition to the highly ionized UV absorption lines 
of NV, CIV and SiIV, also strong ionized absorption is detected in the soft X-
ray spectrum. Since both the UV and X-ray ionized absorber are photoionized in 
QSOs (Weymann 1994, Netzer 1993, Ross \& Fabian 1993),we will present in this 
section, photoionization calculations using 90 version CLOUDY (Ferland 1997). 
The results of these calculations will be used to constrain the physical 
conditions of the X-ray absorbing material. Solar abundances are assumed 
throughout although some earlier evidence suggests the existence of a possible 
heavy element overabundance in the BAL gas(Turnshek 1995).
CLOUDY90 uses a plane parallel slab geometry for the region. 

Mathur et al. (1994) have shown that one should use the actually observed SED of 
an AGN in the application of photoionization codes, rather than a typical quasar 
SED.  Therefore, we take the observed infrared to X-ray spectrum of  
PG 1126-041 as shown in Figure 6, corrected for Galactic reddening of $E(B-
V)=0.08$ in the optical and UV and for the ionized absorption in the X-ray 
 band. The EUV spectrum is a linear interpolation 
from the far UV to the soft X-rays in $log(f_\nu)$ versus $log(\nu)$ space.

To predict the UV absorption lines from the parameters of the warm absorber obtained in \S3, we have calculated a series of models for gas ionized by the  
continuum of PG 1126-041. First, the ionization parameter $\xi$ is converted 
to a dimensionless {\it U= (density of photons at E$>13.6$ eV)/( hydrogen density 
)}. For $\alpha_{uvx}=1.66$, we find $U=0.0427\xi$. The models cover the 
parameter range for the warm absorber, indicated by the probability distribution 
shown in Figure 5 of  $log U=0.26-0.48$, $log N_H = 22.38-22.61$ and 
$n_H=10^9~cm^{-3}$.  The resulting predicted logarithmic column densities (in cm$^{-2}$) are 16.11$^{+0.35}_{-0.29}$, 15.23$^{+1.00}_{-0.95}$, 
15.95$_{-0.72}^{+0.61}$, $<12.00$ for the HI, CIV, NV and SiIV respectively. 
The uncertainties correspond to the minimum and maximum taken for the above 
parameter range.  
 Although the models reproduce the observed CIV and NV column densities 
correctly ({\it cf.} Table~2), the Ly$\alpha$ absorption is far too strong. 
Under the optically thin assumption, the above HI column density corresponds to 
a $W_\lambda(Ly\alpha)=72$\AA, which is a factor of 20 stronger than observed. 
Since the predicted HI column density is at least a factor of two larger than that 
of NV, in the optically thick case the $W_\lambda(Ly\alpha)$ should be similar 
to that of $W_\lambda(NV)$. Such strong absorption in the Ly$\alpha$ is  clearly 
not present in the data. Also, the model predicts that SiIV absorption should 
not be observable and $N(NV)/N(CIV)>2.0$, which would  
imply either, that NV is partially saturated, or that the abundances of C and N 
are different from the solar value.  
  
If the UV absorption lines are produced by an additional gas component, 
the weakness of the Ly$\alpha$ absorption line can be used to constrain the 
turbulent velocity within the absorber. By requiring that the $W_\lambda$ 
produced by the warm absorber material should not be larger than the observed 
value,  
we can derive an upper limit on the turbulent velocity ({\em b}) of the 
absorbing material. The curve-of-growth of the Ly$\alpha$ absorption line shown 
in Figure 7 illustrates that, in order to be consistent with $W_\lambda 
<3.4$\AA, the 
turbulent velocity within the warm absorber must be less than 190 km~s$^{-1}$. 
This limit is far in excess of the expected thermal velocity of ions. However,  
a large velocity gradient within the absorber is predicted in some models, which 
identify the warm absorber with an accretion disk wind. 
Since warm absorption features are dominated by the absorption edges of metal 
ions, the required total hydrogen column density will be less if the metal 
abundances are much higher than the solar value, but this gives only a marginal 
relaxation of the constraint on the turbulent velocity. 
For example, if the HI column density is lowered by a factor of 10, i.e. the 
metal abundances are increased by a factor of $\sim$10, the upper limit of 
{\em b} will be increased by a factor of two at most,  to 300 km~s$^{-1}$ (see 
Figure 7). 

The soft X-ray spectral fit indicates additional absorption due
to cold (neutral or low ionized) material, but little constraints on
the ionization state can be obtained from the X-ray data.
It is likely that part of the UV absorption line is produced by this relatively 
cold material. 
It appears that low column, cold material is located 
outside the warm absorber and thus the ionizing continuum passes through 
the warm
absorber before striking this material.
To model this component, we use the above continuum transmitted
through the warm absorbing material inferred from the X-ray spectral fitting
in the last section. We assume a
particle density $n=10^9$ cm$^{-3}$, and the calculations stop at a column
density of $3\times 10^{20}$cm$^{-2}$. 
The column densities for a few UV absorbing ions are shown
in figure 7. It is clear that this model still predicts too strong Ly$\alpha$ 
absorption. For a wide range of ionization parameters, the model 
predicts $N(HI)\approx N(C IV)$. Therefore, the C and N must be a few times 
over-abundant in the absorbing material in order to explain the weak 
Ly$\alpha$ absorption. 
Moreover, for an ionization parameter range in which 
$N(CIV)\simeq N(NV)$, the SiIV column density is very low, and 
the observed $W_\lambda$ implies that Si should also have a high abundance. 
This situation is similar to that  found for BAL QSOs
(e.g., Junkkarinen, Burbidge \& Smith 1987, Hamann 1998).

The change in the absorption equivalent widths between two observations could be 
due to the fact that either the ionization of the UV absorber is higher in 1995 or, 
if there is a non-radial velocity component and the structure is not symmetric,
the absorber moves out of the line of sight. 
 
Using the scaling law for the BLR size, $R\propto L^{0.5}$, and taking 
NGC5548 as a calibrator, the BLR size in PG 1126-041 is estimated to be 
$\sim 10^{17} $ cm. Since the UV absorber is outside of the BLR, 
the minimal radius can be set to $R_{abs}>10^{17}$ cm. If the absorber is an 
expanding shell of cloudlets with a radial velocity $v_r = 2000$ km/s 
( see Table 3), the distance of the absorber to the center has 
increased by $v_rt\sim 1.5\times 10^{16}$~cm, which is about 
0.1~$R_{abs}$, from 1992 to 1995.
As the absorber is likely in pressure balance with the hot medium, over such 
small distances the decrease of the pressure should be small. Therefore,  
the density of the absorber is similar for the two epochs, and we estimate 
that the ionization parameter in 1995 is about a factor of two higher than 
in 1992. 
A factor of two increase in {\it U} is also expected if the absorbing material 
is in steady 
outflow from the accretion disk.  
The column density of an ion is proportional to the square of the EW of the 
absorption line if the line is severely saturated, and proportional to the 
EW in the optically thin case. As seen in Figure 8, in the range of 
$N(CIV)\simeq N(NV)$, an increase of u by a factor of two will lead to a factor of 2-3
drop in the N V column density and  3-4 in the CIV column density.   
This suggest that 
the absorption lines may  not be severely saturated in the spectrum obtained 
in 1992. 
   
\subsection {Broad Emission Lines}

In order to see if the ionizing continuum illuminating the BELR also
passes through the ionized absorption region, we have computed the 
emission line spectra for two incident ionizing continua, with and 
without filtered by  warm absorption. 
A typical column density of $10^{23}$~cm$^{-2}$ and a
particle density of $10^{10}$~cm$^{-3}$ are adopted. In Figure 9, we plot 
the ratios of line emissivities (the line flux per unit surface area on the 
cloud surface) 
generated by these two input ionizing continua for a wide range of the 
ionization 
parameter. The figure shows that the emissivity of most lines is lower if 
the continuum is absorbed by  warm material. 
This simply reflects the fact that the heating rate decreases due to the 
absorption of soft X-ray photons by the warm absorber.  
The drop is particularly large for O VI and NV lines, whose emission 
strongly depends on the soft X-ray spectrum.  
For these two lines, the suppression 
of NV$\lambda$1240 and OVI$\lambda$1032 relative to Ly$\alpha$ is a factor 
of four and $>$10 for a typical BLR ionization parameter U$\simeq$1. This 
supplies a very sensitive test of whether the ionizing continuum passes through the 
warm absorber before striking the BLR clouds if precise measurement of
more emission lines are available. However, OVI is not observed 
by  IUE and the NV emission is sensitive to the abundance as well. 
Nevertheless, future more sensitive observations will enable us to perform 
such analyses.   

\section{Conclusion}
We have analyzed the UV and X-ray spectra of the X-ray weak AGN PG 1126-041 
and found:\\
(1) The UV spectra show strong highly ionized absorption lines in NV, CIV 
and SiIV,
which extend over a velocity range from $-1000$ to more than $-5000$ km~s$^{-1}$
from the corresponding emission line center. The absorption
column density ratios for metal lines are  similar to those found
in BAL QSOs. The equivalent widths of CIV and NV absorption lines have 
changed by a factor of two over a period of 2.5 yrs. \\
(2) The UV absorption material covers the broad 
emission line region at least partially. \\
(3) Strong ionized absorption has been detected in the soft X-ray
spectrum obtained by {\it ROSAT} simultaneously with the IUE observation in 1992. After 
correcting for absorption, the
optical to X-ray spectral slope is  typical for a radio quiet QSO. \\
(4) The far UV spectrum is steep with $\alpha_{uv}\simeq$ 1.82 and 1.46 
for the 1992 and 1995 data, respectively. Based on the emission line ratios and 
the broad band spectral energy distribution, we argue that the steepness 
of UV spectrum is not due to intrinsic reddening.

From detailed photoionization calculations, we  find:  (1) A single zone warm 
absorber cannot produce the observed equivalent widths of the UV absorption 
lines if the metal abundances are solar. Furthermore, 
in order to be consistent with the observed equivalent width of the $Ly\alpha$ 
absorption line, the turbulent velocity 
of the warm absorber must be less than 190 km s$^{-1}$, this 
imposes serious constraints on a disk wind model.  
(2) The UV absorption lines and their variability cannot be explained by a 
single zone model with solar abundance,
 and the large variability in the 
absorption lines suggest that CIV and NV absorption lines in the 1992 spectrum 
are not 
seriously saturated. (3) The presence of a warm absorber strongly affects the 
ionizing continuum and consequently, has a large impact on the emission line 
spectrum.
\section*{Acknowledgments}
We wish to thank H. Netzer for helpful discussions.
TW and WB are grateful to Masuru 
Matsuoka for his kind hospitality in RIKEN, where part of this work 
was done. We thank the referee for useful comments. 
The CLOUDY code was kindly
supplied by Gary Ferland. TW is supported by the Chinese
Natural Science Foundation and Pandeng project.

\label{lastpage}

\clearpage
\begin{table*}
\begin{minipage}{100mm}
\caption{UV continuum and emission line parameters}
\label{table:uv_spec}
\begin{tabular}{lllll}\hline
  & SWP44822 & SWP44823 & average 1992 & SWP 53285\\ \hline
\multicolumn{5}{l}{\sl continuum:$^{\rm (a)}$} \\
$\beta$  & $0.04\pm 0.03$ & $-0.18\pm 0.02$ & $-0.12\pm 0.06$ & -0.54$\pm$0.01\\ 
A & 1.78$\pm$0.03 & 2.02$\pm$0.02 & 1.90$\pm$0.12 & 4.04$\pm$0.03 \\ 
\multicolumn{5}{l}{\sl line flux:$^{\rm (b)}$} \\
Ly$\alpha$ & 211$\pm$19 & 208$\pm$18 & 210$\pm$5 & 317$\pm$17\\
NV & 73$\pm$9 & 64$\pm$7 & 69$\pm$5 & 67$\pm$10  \\
SiIV+OIV] & 40$\pm$6 & 37$\pm$5 & 39$\pm$3 & 51$\pm$11\\
CIV & 144$\pm$19 & 162$\pm$14 & 153$\pm$9 & 142$\pm$10 \\
HeII & 82$\pm$7 & 39$\pm$4 & 60$\pm$25 & 35$\pm$5 \\
CIII]& & & 51$\pm$16 & \\
MgII & & & 29$\pm$2 & \\ \hline
\end{tabular}

(a) $F_\lambda = A~(\lambda/1000\AA)^\beta$~in units of 10$^{-14}$~erg~cm$^{-
2}$~s$^{-1}$~\AA$^{-1}$.\\
(b) in units of 10$^{-14}$~erg~cm$^{-2}$~s$^{-1}$.
\end{minipage}
\end{table*}

\begin{table*}
\begin{minipage}{120mm}
\caption{absorption line parameters}
\label{table:uv_abs}
\begin{tabular}{lrrrrr} \\
\hline
 & SWP44822 &SWP44823 & 1992 average & N$_{1992}$ & SWP 53285  \\
\hline
\multicolumn{6}{l}{central velocity and width (km/s)} \\
V          & 1826$\pm$70 & 1990$\pm$70 & 1910$\pm$80 & & 2200$\pm$70 \\
$\sigma_a$ & 3155$\pm$157 & 2875$\pm$115 & 3000$\pm$150& & 2010$\pm$50 \\
\multicolumn{5}{l}{equivalent widths (\AA)} \\
Ly$\alpha$ & 0.7$\pm$0.9 & 2.7$\pm$0.7  & 1.7$\pm$1.0 & 1.6~10$^{14}$ &
3.3$\pm$0.5 \\ 
NV$\lambda$1240& 13.9$\pm$0.8 &  11.1$\pm$0.6  & 12.5$\pm$1.4 &
1.6~10$^{15}$ &
6.8$\pm$0.5\\  
SiIV$\lambda$1397 & 5.7$\pm$1.7 & 6.3$\pm$1.3 & 6.0$\pm$0.3 &2.3~$10^{14}$
& 0.2
$\pm$0.4\\ 
CIV$\lambda$1549 & 14.3$\pm$1.0 & 15.4$\pm$0.7 & 14.9$\pm$0.5
&1.2~$10^{15}$ & 5
.3$\pm$0.7\\ \hline  
\end{tabular}

N$_{1992}$ is the column density of the ion under optically thin conditions
(in units of $cm^{-2}$).
\end{minipage}
\end{table*}

\begin{table*}
\begin{minipage}{150mm}
\caption{ROSAT-PSPC spectral fitting}
\label{table:rosat_spec}
\begin{tabular}{llllllll} \hline
model & $N_c	^{\rm (a)}$   & $\Gamma$ & N$_{1keV}^{\rm (b)}$ & $E_{edge}$ 
or  $\xi$& $\tau$ or $N_w^{\rm (a)}$ & $\chi^2_{\nu, red}$/dof \\ \hline
PL   & 4.4 ({\it f})    & 3.20 & 0.3 &   &      & 4.72/11 \\
PL   & 0.5    & 1.48 & 0.37 &   &     &  4.00/10\\
Edge & 5.6$^{+1.0}_{-0.9}$ & 2.2$_{-0.3}^{+0.3}$ & 2.2$\pm$0.3 & 
0.56$\pm$0.03& 15.4$_{-6.0}^{(c)}$ & 1.01/8 \\
WAB  & 7.0$_{-0.9}^{+1.0}$ & 2.8$_{-1.8}^{+1.3}$ & 6.9$\pm$2.5 & 55$_{-
12}^{+16}$ & 320$_{-80}^{+90}$ & 1.23/8 \\ \hline
\end{tabular}

 (a) In units of 10$^{20}$~cm$^{-2}$. \\
 (b) In units of 10$^{-4}$~cm$^{-2}$~s$^{-1}$~keV$^{-1}$. \\
 (c) The upper bound of the parameter was 20.
\end{minipage}
\end{table*}

\begin{figure*}
\vspace{1cm}
\caption{Observed Ultraviolet Spectrum of PG1126$-$041 processed
by NEWSIPS and corrected for Galactic reddening of E(B$-$V)=0.08 (see text). NEWSIPS 
errors are plotted as dotted lines. The major emission and absorption lines are 
indicated. The reality of the second peak in the CIV profile could not be unambiguously established although some structure appears to be present in both 1992 spectra. Upon inspection of the SILO file, it could very well be associated with a camera hotspot on the spectrum in SWP44823.}
\caption {A detailed view of emission line and the associated absorption
line profiles in velocity space. A redshift z=0.060 is adopted. The emission lines do not have their peaks at the same velocity (see also text). The 1992 
spectrum is shown as thin line and that of 1995 as thick line. The spectra are normalized to the local continuum level. Note that in 1995, when the continuum level was twice that of 1992 (see also table~1) the absorption lines were considerably weaker.}
\caption{Fitted SWP spectra of 1992 and 1995 as described in the text.}
\caption{Fits of the {\it ROSAT} soft X-ray spectrum with a single 
power-law model (dotted line), with an additional absorption edge (dotted-dash line) 
and with an ionized warm absorption model (dash line). The top panel shows the unfolded spectrum, with the model fits as indicated. The lower panels show the
residuals for a single power-law model (PL); with absorption edge (EDGE); and with a warm absorber model (WARM). See text for a detailed description of the models.}
\caption{60, 90 and 95 per cent contours (upper limits for two interesting 
parameters) for the ionization parameter against the column density for 
the warm absorption model.} 
\caption{Broadband spectral energy distribution of PG1126-041. The UV data have 
been taken during the ROSAT observation and can be 
considered as essentially simultaneous. The full line in the X-rays
represents the derived input spectrum after correcting the observed X-ray  
spectrum (also shown) for Galactic absorption  and  
 for the presence of a warm absorber (see section 3).}
\caption{The growth curve for the Ly$\alpha$ absorption line. The solid line 
corresponds to a column density of HI of 1.3$\times 10^{16}$~cm$^{-2}$, the 
value predicted by the warm absorption model with cosmic abundances, the dashed 
line for an N(HI)=1.3$\times 10^{15}$~cm$^{-2}$, approximating the case of a 
warm absorption model 
with metal abundances, a factor of 10 times solar values.} 
\caption{Predicted ion column densities for major ultraviolet absorption
ions for a hydrogen column density of 3$\times 10^{20}$~cm$^{-2}$ (see text 
for details).}
\caption {Plot of line emissivity ratios against photoionization parameter
for a particle density n$_H$=10$^{10}$~cm$^{-3}$ and column density
N$_H$=10$^{23}$~cm$^{-3}$. The {\it I$_w$} is the line intensity calculated 
assuming that the ionizing continuum passes through the warm absorber before 
striking the BLR. }
\end{figure*}

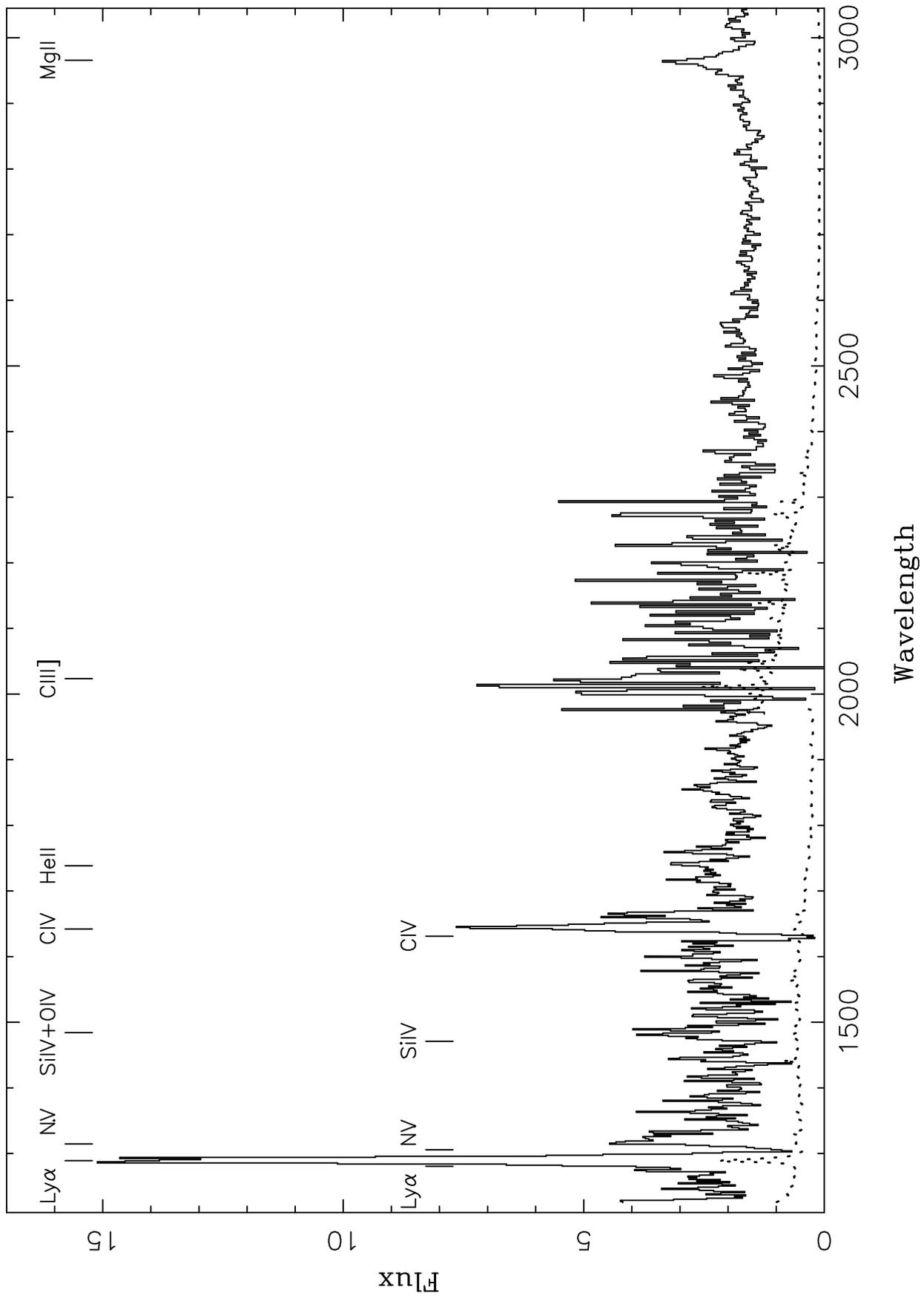
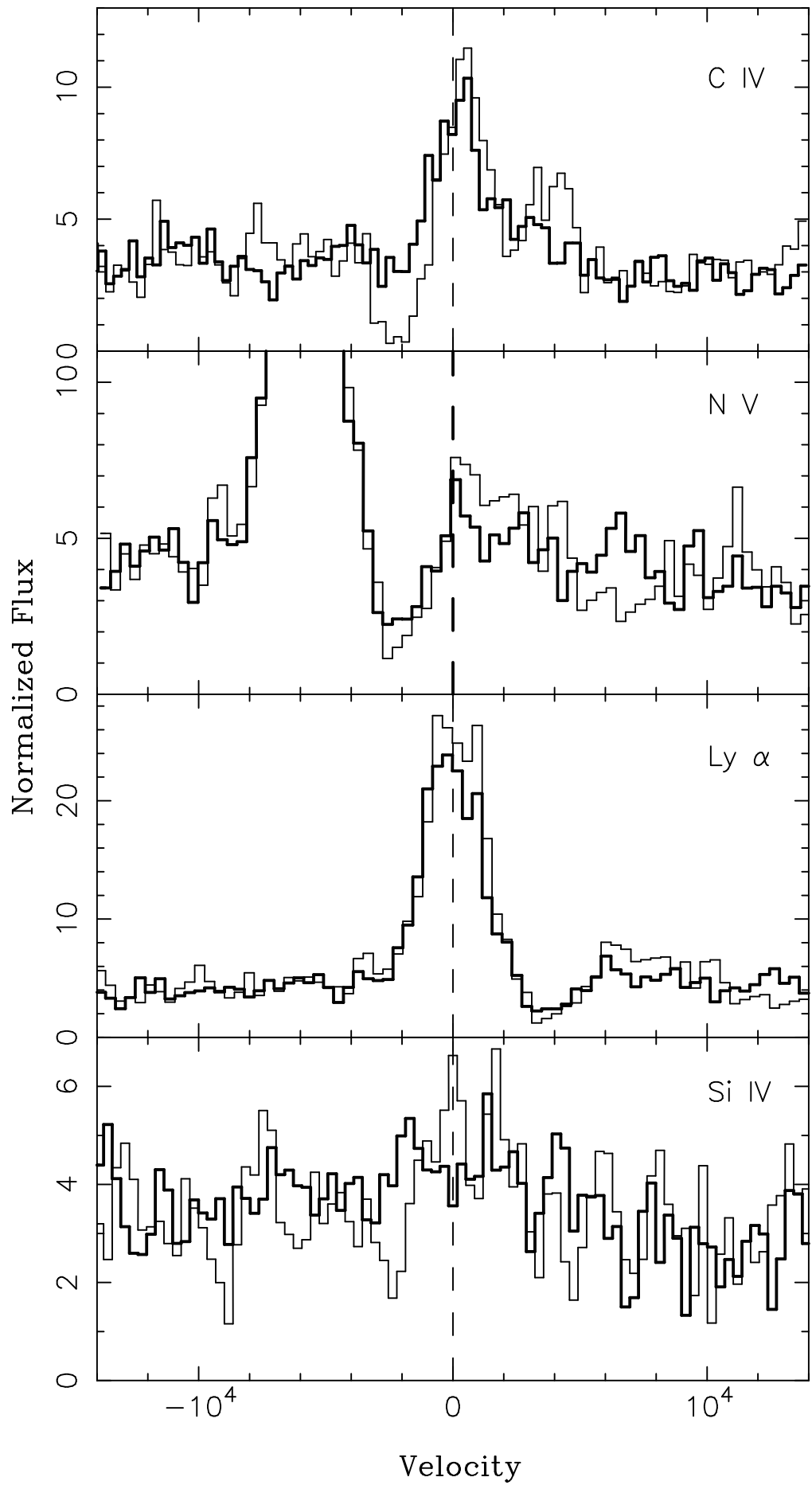
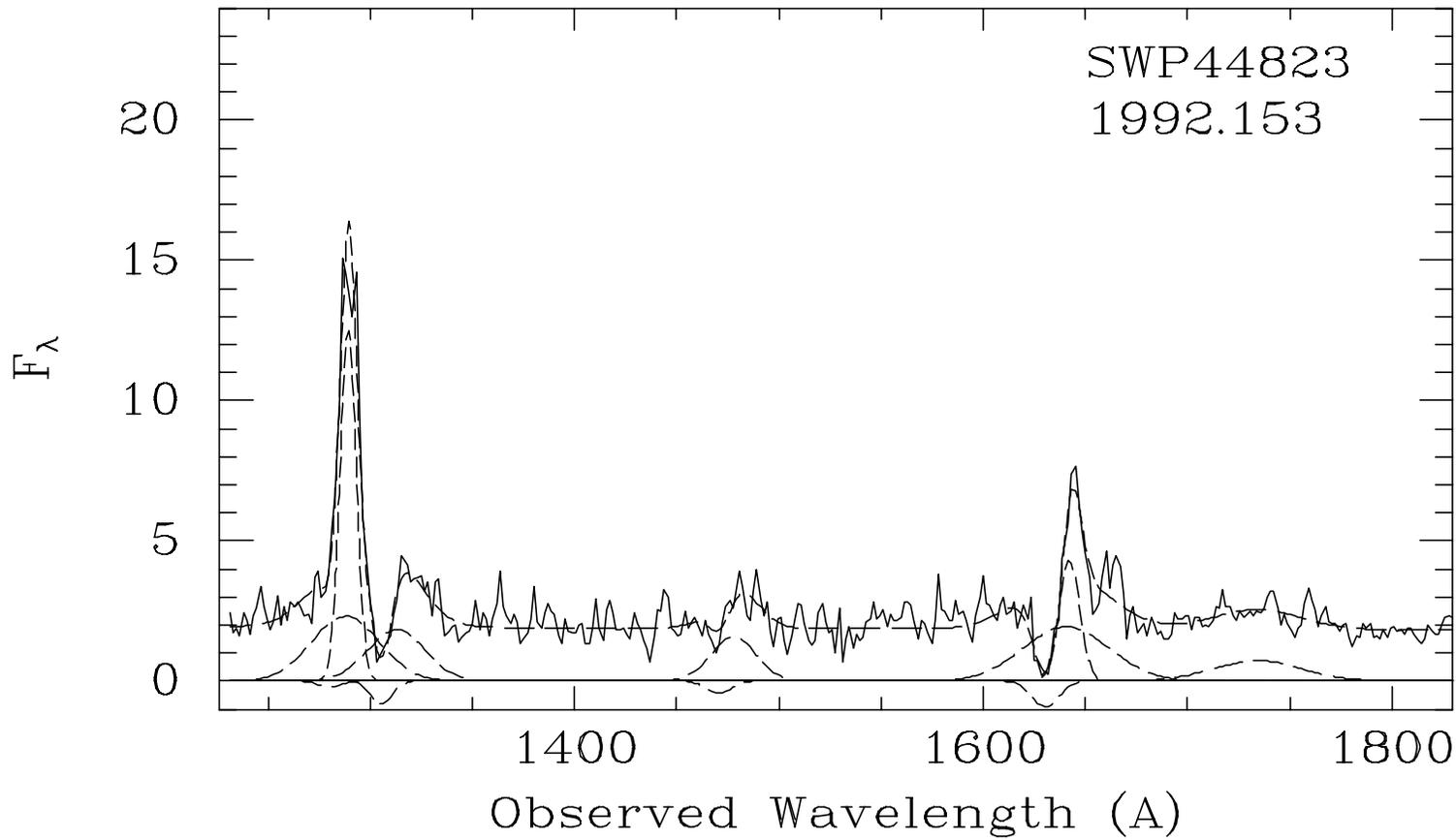
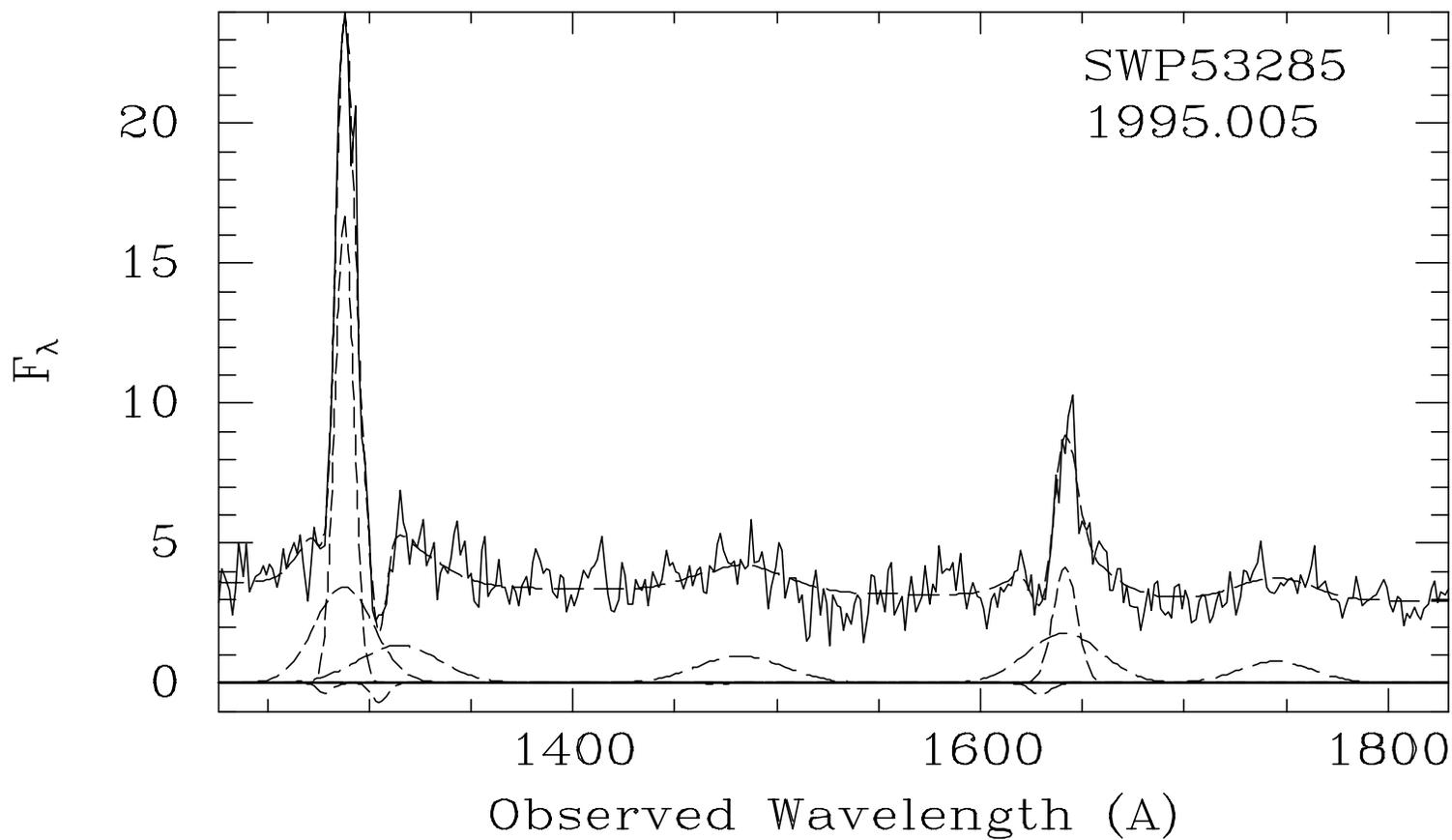
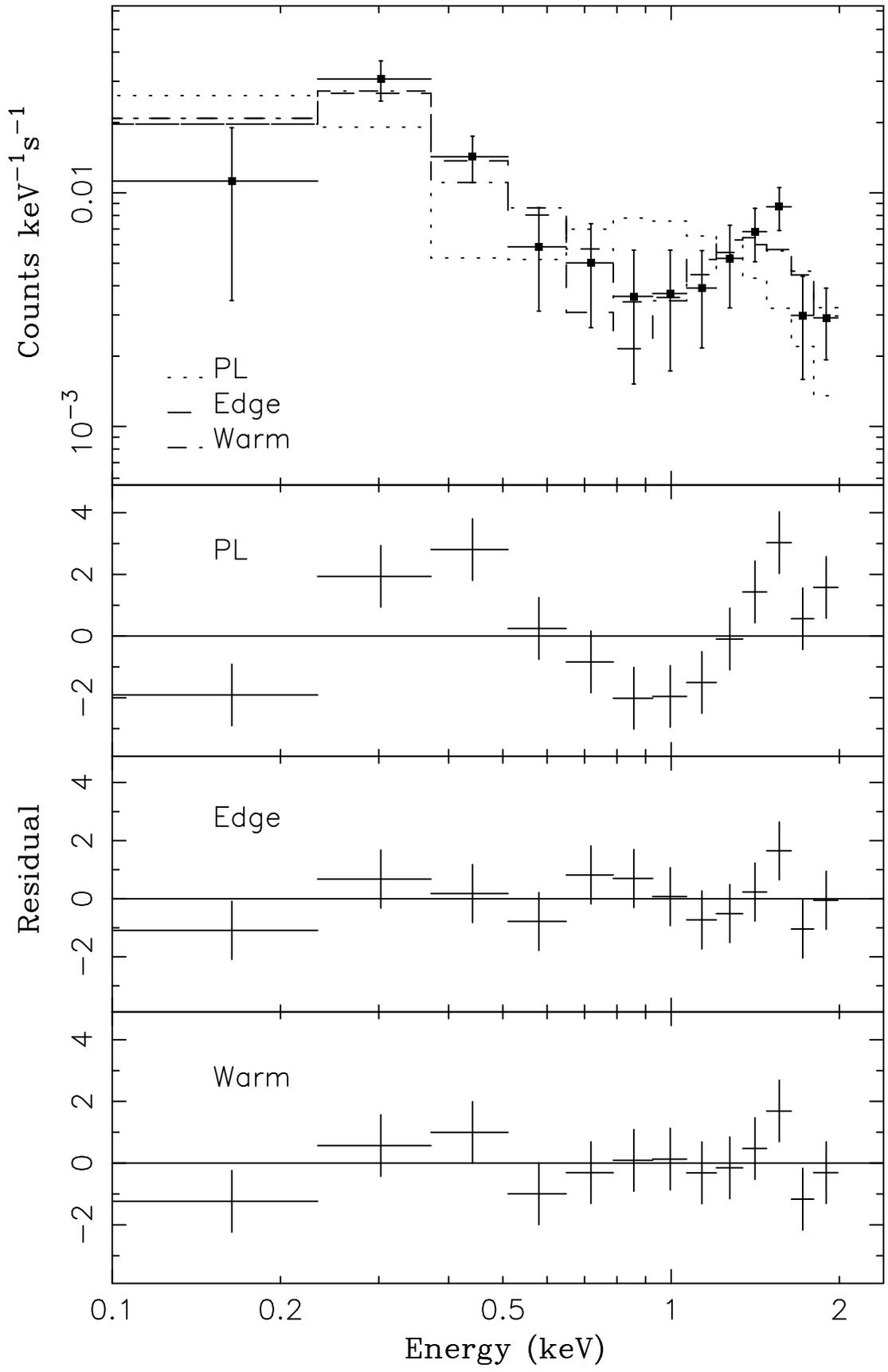
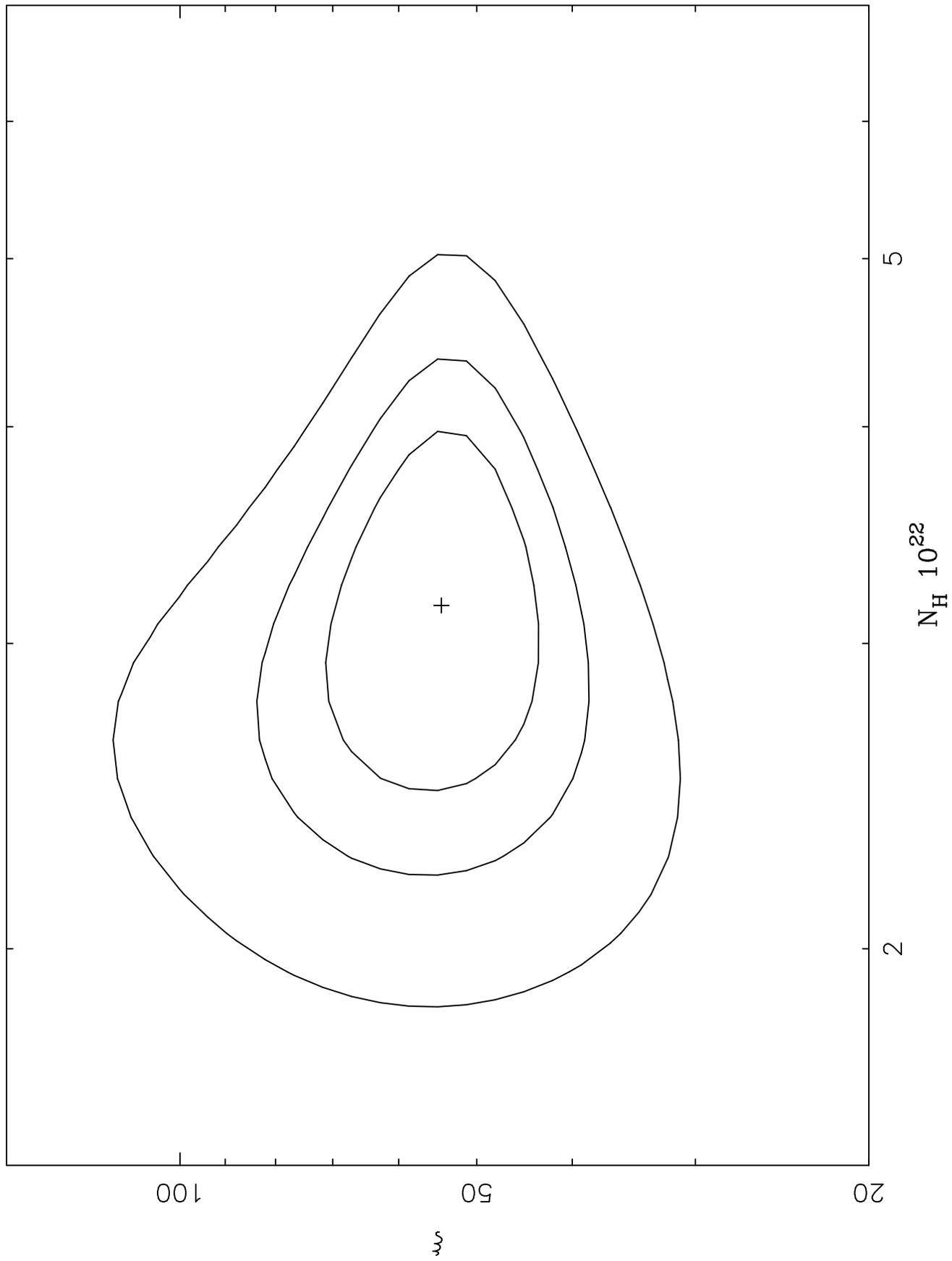
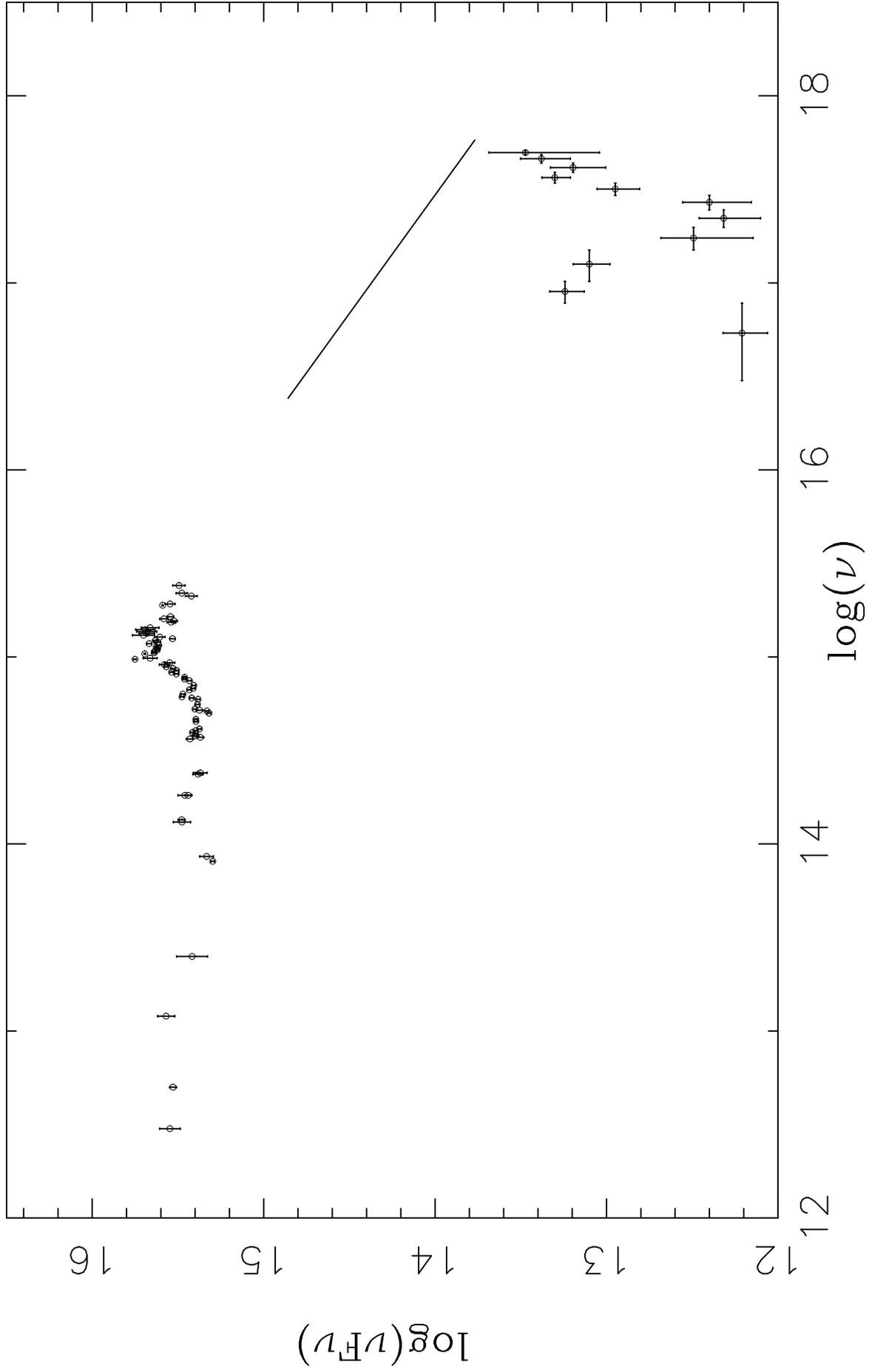
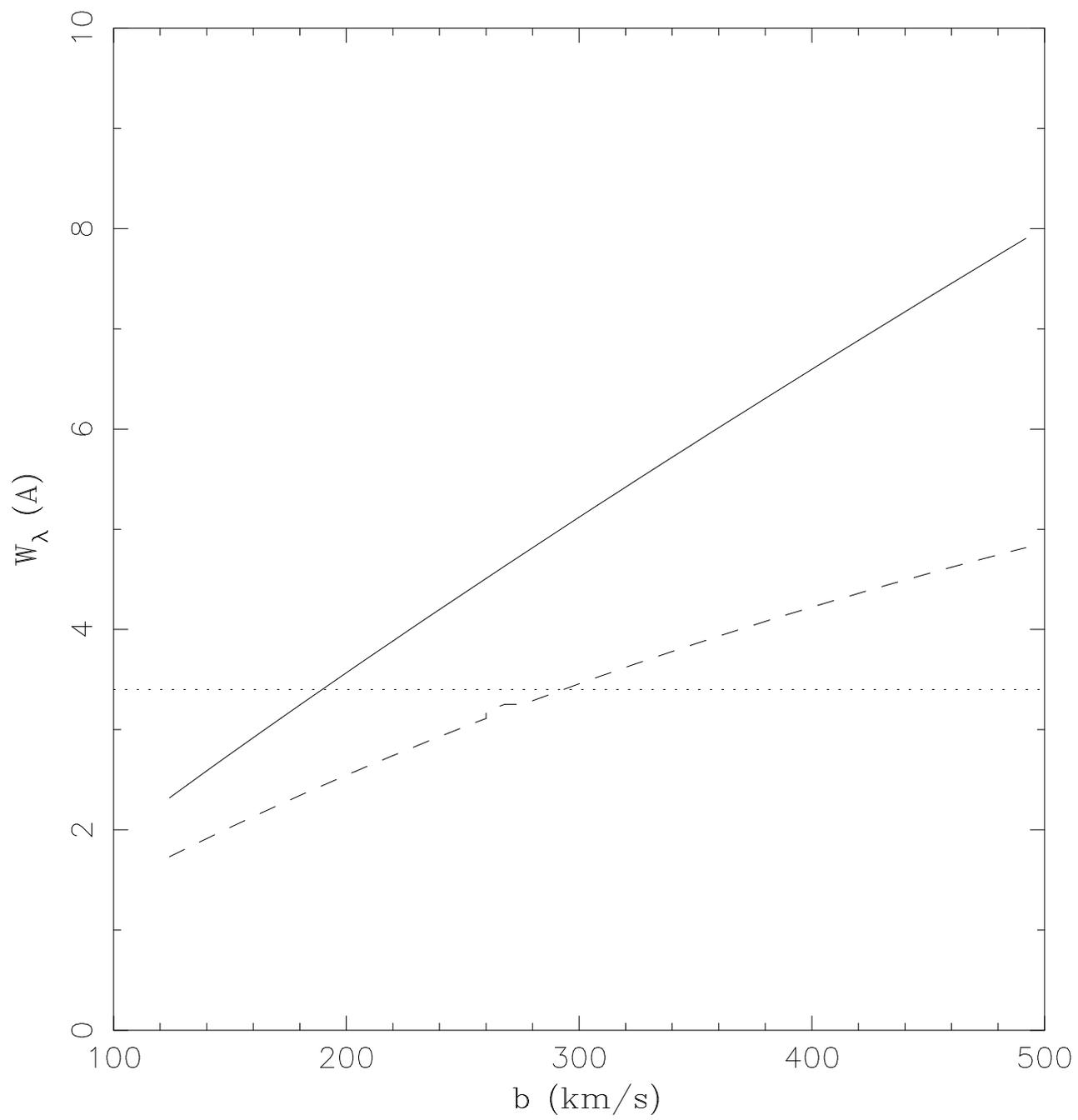
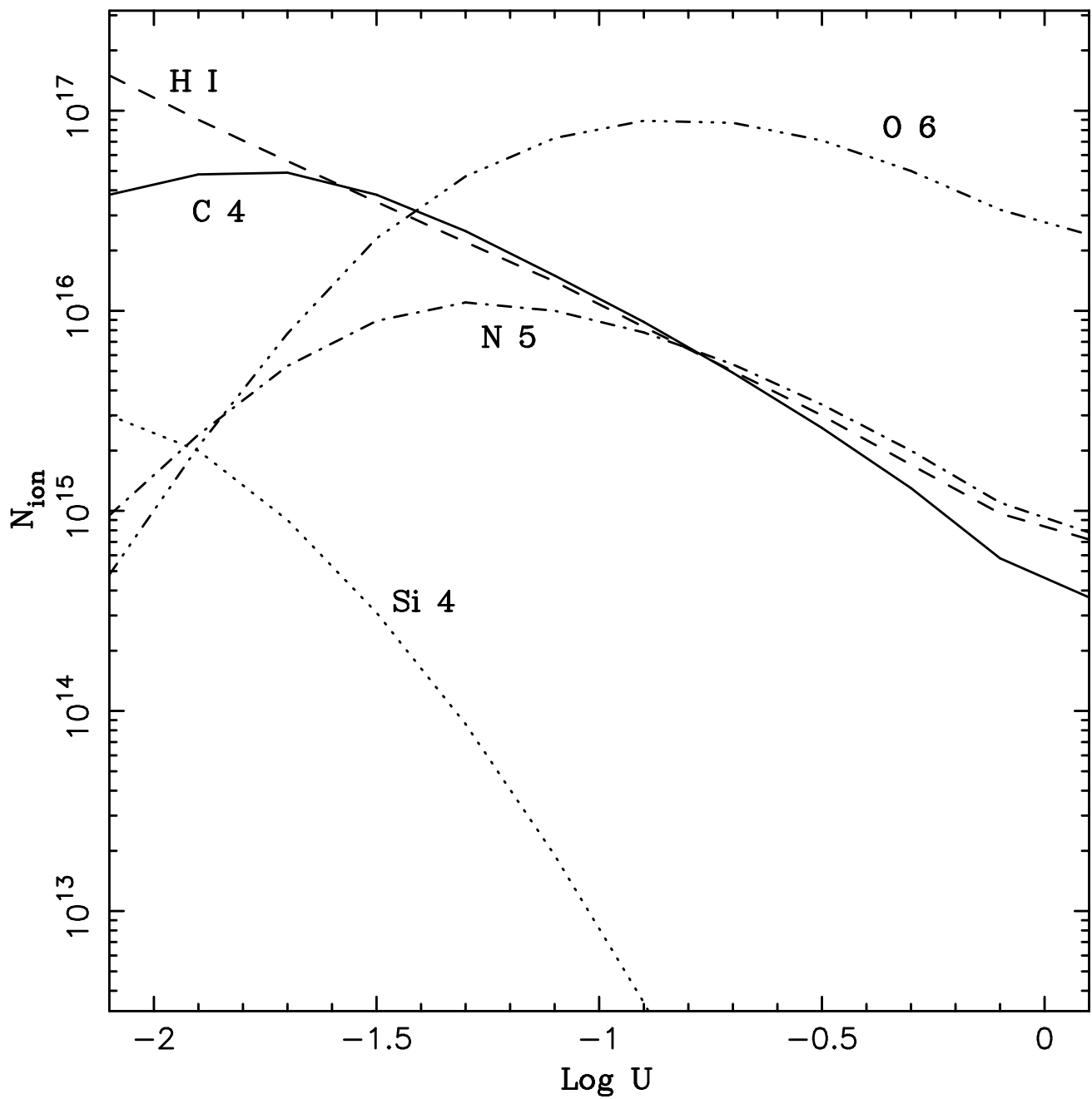
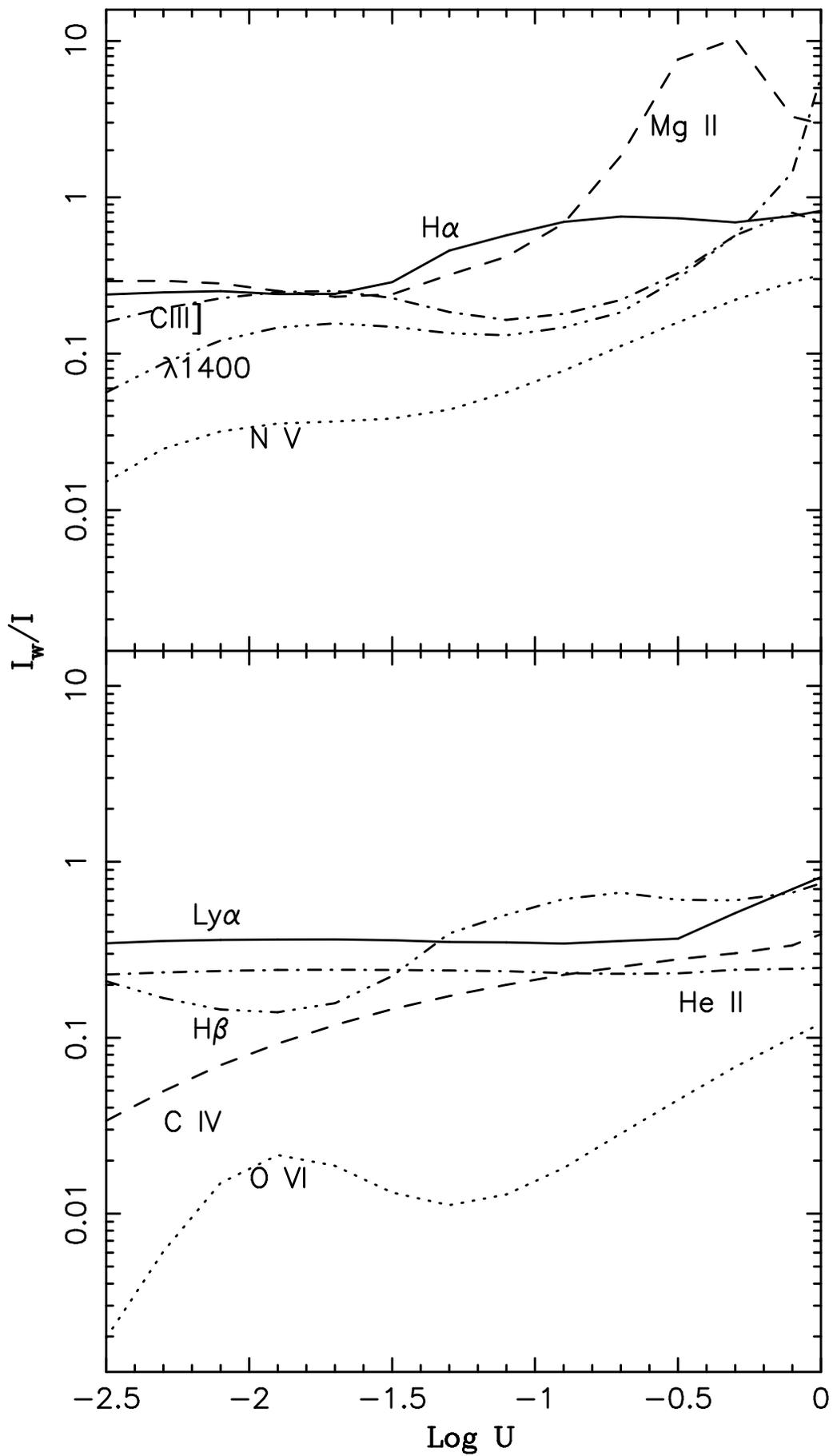
\end{document}